\documentclass{nature}

\usepackage{amsmath,amssymb}
\usepackage{graphicx}
\usepackage{color}

\title{Detection of Gravitational Wave Emission by Supermassive Black Hole Binaries Through Tidal Disruption Flares}

\author{}
\author{Kimitake Hayasaki$^{1,2\ast}$, Abraham Loeb$^{2}$\\
$^\ast$
E-mail:  kimi@cbnu.ac.kr
}
\begin{document}

\maketitle

\begin{affiliations}
 \item Department of Astronomy and Space Science, Chungbuk National University, Cheongju 361-763, Korea
\item Harvard-Smithsonian Center for Astrophysics, 60 Garden Street, Cambridge, MA\,02138, USA
\end{affiliations}

\begin{abstract}
Galaxy mergers produce supermassive black hole binaries, which 
emit gravitational waves prior to their coalescence. We perform 
three-dimensional hydrodynamic simulations to study the tidal disruption 
of stars by such a binary in the final centuries of its life. We find that the 
gas stream of the stellar debris moves chaotically in the binary potential 
and forms accretion disks around both black holes. 
The accretion light curve is modulated over the binary orbital period owing to 
relativistic beaming. This periodic signal allows to detect the decay of the binary 
orbit due to gravitational wave emission by observing two tidal disruption events 
that are separated by more than a decade.
\end{abstract}

%
\section*{Introduction}
%

%
%
Most galaxies are inferred to harbor supermassive black holes (SMBHs) with 
masses in the range of $10^{5-9}\,{\rm M}_\odot$ at their centers 
\cite{kr95,kho13}, based on observations of stellar proper motion \cite{rs+02}, 
stellar velocity dispersion \cite{mtr+98} or accretion luminosity \cite{mm+95}. 
Most of time, their nuclei do not exhibit significant activities \cite{rg+03}, 
but in a small fraction of all galaxies the inflow of cold gas triggers active galactic 
nuclei (AGNs) which produce intense radiation and powerful outflows. 
Tidal disruption events (TDEs) provide a distinct opportunity to probe dormant 
SMBHs in inactive galaxies (see a recent general review \cite{komo15}).


%
%
SMBH binaries with a sub-parsec ($\lesssim1\,\rm{pc}$) separation are considered 
to be the end product of a galaxy merger before two black holes finally coalesce \cite{sill88,ghm15}. The merger of two SMBHs goes through the following processes \cite{bege80}: first, the SMBHs sink towards each other via dynamical friction on the surrounding stars and gas. Once their separation drops below a parsec, a hard binary forms \cite{may07} and tightens further due to other mechanisms\cite{iv99,kh09} until gravitational wave emission takes over at the final stage\cite{cd11}. 

%
%
The observed TDE rate on single SMBHs is $\sim 10^{-5} \rm{yr}^{-1}$ per galaxy \cite{dbeb02,vf14}, 
although another recent survey suggests that the rate could be 
one-order of magnitude higher \cite{holo+15}. This higher rate has been implied by 
the recent theoretical works based on two-body relaxation \cite{wm04,nb14}.
However, SMBH binaries are expected to exhibit an enhanced TDE rate of up to $10^{-1}~{\rm yr}^{-1}$ from chaotic orbital evolution and the Kozai-Lidov effect\cite{ips05,cms09,li15}. 
Recent work suggests that multiple TDEs may occur in merging SMBH binaries with a rate of up to a few times over an observing period of five years with the {\it Large Synoptic Survey Telescope} (LSST, http://www.lsst.org)\cite{wb11}. It is less clear what characteristic flares such TDEs would exhibit, although interruptions in the light curve are likely if the orbital period of SMBH binaries is less than the period of an observable emission \cite{llc09,llk14,rnd15}.

%
%
%
%
If the binding energy of the approaching star relative to either component
of the binary is negative inside the tidal disruption radius, the star should be 
finally tidally disrupted. Here, we study how stars migrating through the binary 
Lagrange points are tidally disrupted by the SMBHs. In Section~2, we describe 
the numerical method used in our simulations of these TDEs. 
We then analyze our numerical results in Section~3. 
Finally, we summarize our main conclusions in Section~4.

%
\section*{Methods}
\label{sec:sph-code}
%

%
%
\subsection{Test-particle limit.}
%
We consider a system composed of a star orbiting 
around a circular SMBH binary with mass ratio 
$q=M_2/M_1=0.1$. 
The initial position and velocity in our simulation 
are correspondingly the Lagrange point L2 and the Keplerian 
velocity there. This situation would be, for example, 
realized in the case of a massive circumbinary disk \cite{pau13}, 
where the disk makes stars migrate inwards and approach 
the Lagrange points around the binary (L2 or L3) 
via a mass stream from the disk's inner edge \cite{kh07}.
The dynamics of the star can be treated as a restricted 
three-body problem. We solve numerically the motion 
of the star as a test particle in the binary potential.

\begin{figure*}
\resizebox{\hsize}{!}{
\includegraphics[width=10cm]{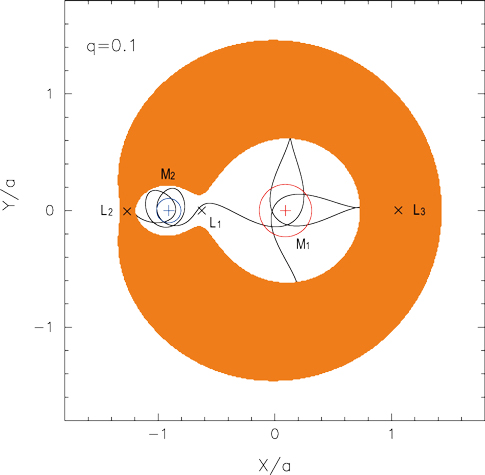}
}
\caption{
Orbit of a test particle in a co-rotating frame around a circular binary with $M_2/M_1=0.1$. The red and blue circles shows the tidal disruption radii around the primary and secondary black holes, respectively. The three black crosses mark the Lagrange points, $L_1$, $L_2$, and $L_3$. Both axes are normalized by the binary semi-major axis. The black solid line shows the motion of a test particle inside the zero-velocity curve for a Jacobi constant $C_{\rm J}=3.45$, and the shaded area outside the curve denotes the forbidden zone \cite{mcd99}.
}
\label{fig:bsmbhs} 
\end{figure*}

%
%
In the test-particle limit, the star is initially located just 
inside the L2 point with a Keplerian velocity, and then 
is trapped by the binary potential and freely moves between 
two black holes inside the inner Roche-lobe of the binary. 
The outside of the inner lobe is marked by the shaded area 
of Figure~\ref{fig:bsmbhs} as the forbidden zone \cite{mcd99}.
The star is tidally disrupted once it enters the tidal disruption radius:
\begin{equation}
r_{t}\simeq \left(\frac{M_{\rm BH}}{m_*}\right)^{1/3}\,r_{*}\sim 24\,
\left(\frac{M_{\rm{BH}}}{10^6\,{\rm M}_\odot}\right)^{-2/3}
\left(\frac{m_*}{{\rm M}_\odot}\right)^{-1/3}
\left(\frac{r_*}{{\rm R}_\odot}\right)
\,r_{\rm S},
\label{eq:rt}
\end{equation}
where $M_{\rm{BH}}$ is the SMBH mass, $m_*$ and $r_*$ are the star's mass and radius, and 
$r_{\rm{S}}=2{\rm G}M_{\rm BH}/{\rm c}^2$ is the Schwarzschild radius.
The corresponding orbit is depicted in Figure~\ref{fig:bsmbhs}. 
The fate of the tidally disrupted star is similar to that of an eccentric 
TDE \cite{hsl13}, in which the TDE accretion rate deviates significantly 
from the standard $t^{-5/3}$ time evolution \cite{bonne15,hsl15}.

%
%
\subsection{SPH simulations.}
%
Our simulations are performed with 
a three-dimensional Smoothed Particle Hydrodynamics (SPH) code \cite{bate95}.
The hydrodynamic equations are integrated using a second-order 
Runge-Kutta-Fehlberg integrator with individual time steps for each 
particle, leading to substantial savings in time for a large dynamic range 
of timescales. We also use a variable smoothing length scheme to find 
the relevant spatial resolution in our code, but ignore the term proportional 
to the gradient of the smoothing length. We adopt standard the SPH artificial 
viscosity parameters $\alpha_\mathrm{SPH}=1$ and $\beta_\mathrm{SPH}=2$.

%
%
The photon diffusion timescale of the stellar debris  
is given by 
\begin{eqnarray}
t_{\rm{diff}}=\frac{H}{c}\tau
\simeq
6.1\times10^8
\left(\frac{\Sigma}{\Sigma_0}\right)
\left(\frac{H}{\Delta{r}}\right)
\left(\frac{r}{r_{\rm{t}}}\right)^{-1}\,{\rm s},
\label{eq:tdiff}
\end{eqnarray}
where $H$ is the debris scale height and $\tau=\kappa_{\rm es}\Sigma\sim2.6\times10^6\,(\Sigma/\Sigma_0)$ is the optical depth for electron scattering
with $\kappa_{\rm{es}}=0.4\,\rm{cm^2\,g^{-1}}$, 
and $\Sigma$ is surface density of the stellar debris with the fiducial value of
\begin{eqnarray}
\Sigma_0\equiv\frac{m_*}{2\pi{r}\Delta{r}}
&\simeq&
6.5\times10^6\,
\left(\frac{m_*}{M_\odot}\right)
\left(\frac{r}{r_{\rm{t}}}\right)^{-1}
\left(\frac{\Delta{r}}{r_{\rm{t}}}\right)^{-1}
\,{\rm{g\,cm^{-2}}},
\label{eq:sigma0}
\end{eqnarray}
where $r$ and $\Delta{r}$ are the radius and width of the debris ring. 
Recent three-dimensional radiation magneto-hydrodynamic 
simulations showed that the magnetic buoyancy induced through the 
magneto-rotational instability (MRI) \cite{bh91} provides an efficient radiative 
cooling mechanism for super-Eddington accretion flows \cite{yanfei14}. 
In this regime, their preliminary simulations indicate that the radiative cooling 
speed is determined by the sound speed of the radiation.
In our simulation, the stellar debris loses its orbital energy and then circularizes via a shock 
into an accretion flow at a super-Eddington rate at $\sim{r}_{\rm t}$. 
The initial radiation energy density is $\sim\rho_0v^2_{\rm t}$, where $\rho_0=\Sigma_0/r_{\rm t}$ and $v_{\rm t}=\sqrt{GM_{\rm BH}/r_{\rm t}}$, and so the radiation sound speed is estimated to be $c_{\rm{s,rad}}\sim\sqrt{P_{\rm rad}/\rho_0}\sim0.08c$. Thus, the resulting cooling timescale $r_{\rm t}/c_{\rm s,rad}$ is estimated to be $\sim2.8\times10^3\,{\rm s}$, shorter than the heating timescale by the shock $2\pi\sqrt{r_{\rm t}^3/GM_{\rm BH}}\sim10^4\,{\rm s}$. 
It takes several dynamical times for the magnetic field to grow via the MRI, after which the efficient cooling reduces the thickness of the debris streams. 
Thus, we use a simple polytropic equation of state: $P=K\rho^{\gamma}$, where $P$ and $\rho$ are the gas pressure and density, $\gamma=5/3$ and 
$K$ is a constant. This ensures that the radiative cooling is efficient throughout our simulation.

%
%
We model the initial star as a polytrope with the polytrope index $n=1.5$ 
in hydrostatic equilibrium. The star is set in motion through the gravitational 
field of the SMBH binary with the following parameters: $m_*=1M_\odot$, 
$r_*=1R_\odot$, $M_{\rm{BH}}=10^6M_\odot$, $a=100r_{\rm S}$, and 
$q=0.1$. The simulations follow $0.5$ million SPH particles for $14$ 
binary orbital periods, where a single period $P_{\rm orb}\sim 1\,\rm{day}$.
%

%
\section*{Results}
\label{sec:btdes}
%
%
\subsection{Tidal disruption of a star.}

\begin{figure*}
\begin{center}
\includegraphics[width=14cm]{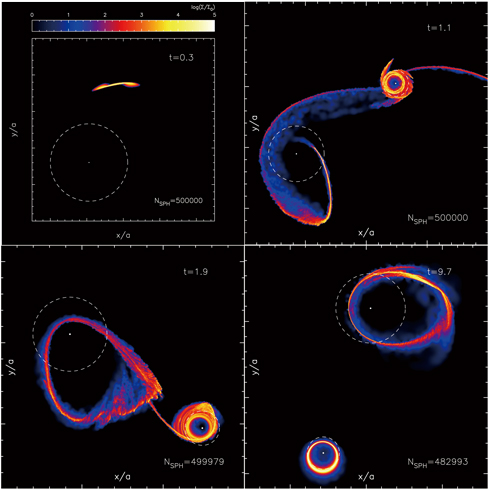}
\end{center}
\caption{
Gas density maps, projected on the orbital (x-y) plane, during a tidal disruption event 
in a circular SMBH binary with $M_{\rm BH}=10^6M_\odot$, $a=100r_{\rm S}$, 
and $q=0.1$. The run time $t$ is in units of $P_{\rm orb}$ as labeled in the 
top-right corner of each panel. The number of SPH particles is indicated at 
the bottom-right corner. The panels are shown in chronological order. The coloration 
indicates the density in five orders of magnitude on a logarithmic scale normalized 
by $\Sigma_0=6.5\times10^6\,\rm{g\,cm^{-2}}$. The dashed circles indicate the tidal 
disruption radii for the primary and secondary SMBHs.
}
\label{fig:snap} 
\end{figure*}

Figure~\ref{fig:snap} shows the evolutionary sequence of density 
maps for the stellar debris of a TDE around the SMBH binary.
A star approaching the binary from the L2 point is tidally disrupted 
by the secondary black hole. Part of the stellar debris is not 
gravitationally bound, while the rest accretes onto the secondary 
SMBH. The unbound debris moves chaotically in the binary potential 
and forms accretion disks around both black holes by transferring 
mass along the potential valley through the L1 point.

\begin{figure*}
\includegraphics[width=9cm]{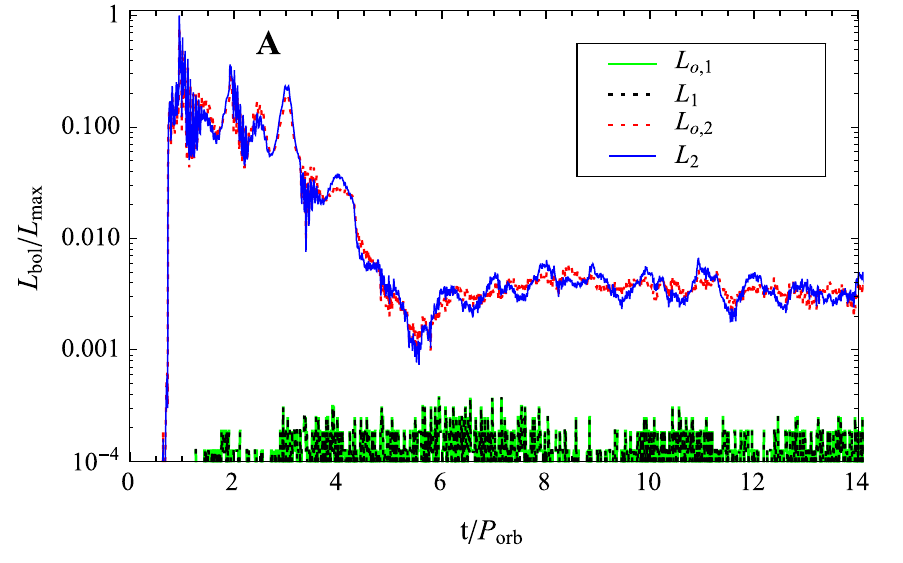}
\includegraphics[width=9cm]{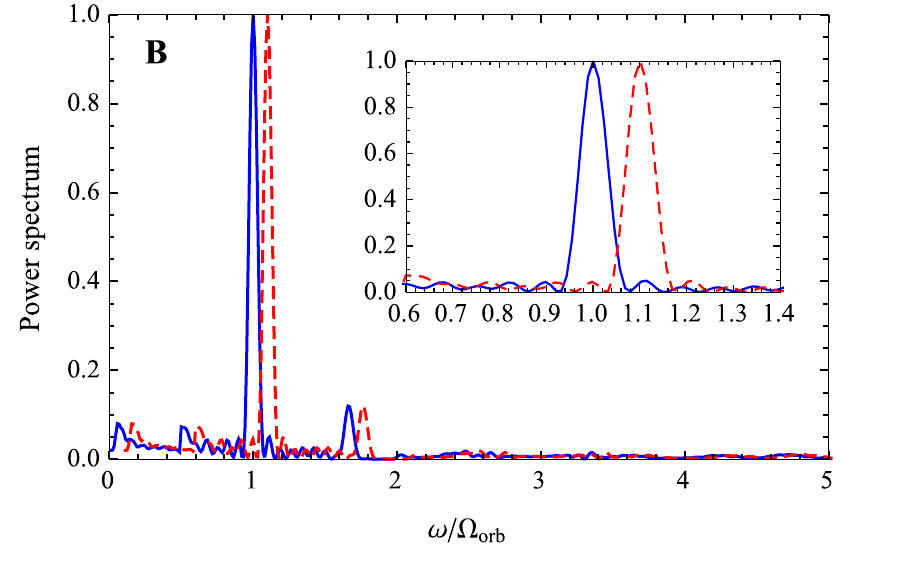}\\
\includegraphics[width=9cm]{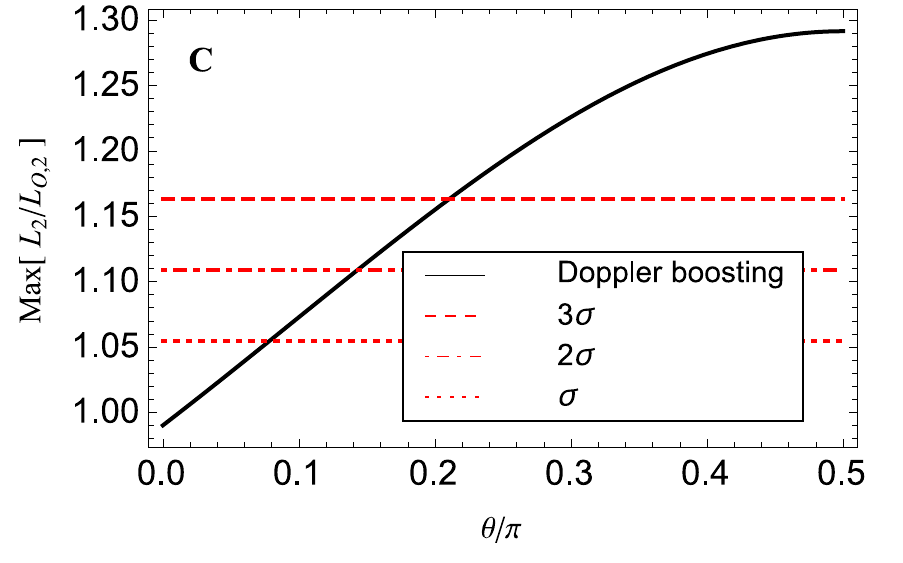}
\caption{
(\textbf{A}) Bolometric luminosity of the primary (black and green) and 
secondary (red and blue) black holes normalized by its maximum 
value $L_{\rm max}=\max\{L_{\rm{o},1}+L_{\rm{o},2}\}$. The black and blue 
lines include relativistic beaming. 
(\textbf{B}) Power Spectrum of the bolometric light curves with relativistic 
beaming as a function of frequency $\omega$ in units of $\Omega_{\rm orb}=
2\pi/P_{\rm orb}$. The blue solid line shows the power spectrum of the first 
TDE, and the red dashed line shows that of the second TDE 50 years later. 
The half maximum full-width are given by $0.05240\pm0.000563$ with a 
fixed peak position $0.99027$. If the two peaks are sampled by $N$ 
data points, it should be possible to separate them even after a time interval 
of $50/\sqrt{N}$ years. (\textbf{C}) Inclination angle dependence of the 
maximum value of $L_2/L_{o,2}$. The dashed, dot-dashed, dotted lines are 
the ones corresponding to 1 to 3 $\sigma$ of the fluctuations in $L_{o,2}$.
}
\label{fig:lc} 
\end{figure*}

%
%
%
The resulting mass accretion rates are shown in Figure~\ref{fig:lc}A.
The green and red dotted lines show the bolometric light curves of the 
primary and secondary black holes, respectively. The luminosities are 
derived as $L_{{\rm{o}},i}\propto\dot{M}_{i}c^2$ with $i=1,2$, where 
$\dot{M}_1$ and $\dot{M}_2$ are the mass accretion rates of the primary 
and secondary black holes based on the simulation. The accretion rates 
are estimated at the radius of the innermost stable circular orbit of a 
Schwarzschild black hole, $3r_{\rm  S}$. The black and blue lines show the 
bolometric light curves, including the relativistic doppler beaming effect due 
to the orbital motion of the binary \cite{dzd15},
\begin{eqnarray}
L_i=L_{{\rm o},i}\left[\frac{1}{\Gamma(1-\beta_i)}\right]^4\,\,\,(i=1,2)
\label{eq:blc}
\end{eqnarray}
where $\Gamma=1/\sqrt{1-(v_{\rm orb}/c)^2}$ and $v_{\rm orb}=\sqrt{GM/a}$ 
are the Lorentz factor and the orbital speed, respectively, and
\begin{eqnarray}
  \left\{
    \begin{array}{l}
       \beta_1=(v_{\rm orb}/c)(q/1+q)\sin\theta\sin(\Omega_{\rm orb}t) \\
       \beta_2=(v_{\rm orb}/c)(1/1+q)\sin\theta\cos(\Omega_{\rm orb}t)
    \end{array}
  \right.
\end{eqnarray}
with $\Omega_{\rm orb}=\sqrt{GM_{\rm BH}/a^3}$ being the orbital frequency and 
$\theta$ being the inclination angle between the line-of-sight and the binary 
orbital plane, which is assumed to be $\pi/2$ in the current simulation. Note that $\dot{M}_{1}$ 
intrinsically shows the bursty nature and is more than one order of magnitude smaller than $\dot{M}_2$, since the angular momentum of the stellar debris relative to the primary black hole is much larger than that of the secondary black hole. In addition, 
the periodicity of the bolometric light curves of the secondary black hole is enhanced 
by relativistic beaming.

%
%
%
Figure~\ref{fig:lc}B shows the power spectrum of the total light curve 
with relativistic beaming. The blue solid and red dashed lines mark the 
simulated and predicted power spectra, respectively. The blue line indicates 
a sharp peak at the binary orbital frequency. Since the full-width at half 
maximum of the peak is $\sim0.052\,\Omega_{\rm orb}$, the shift of the 
peak over time due to orbital decay should be larger than $\pm0.052\,\Omega_{\rm orb}$ 
in order to be easily noticeable. For a fiducial TDE rate of $\sim0.02$ 
per year, a second TDE would likely occur 50 years after the first TDE, 
and the shift in the power spectrum due to the orbital decay by gravitational 
wave emission will be noticeable, as shown by the red dashed line of the figure.
If the peak is sampled by N data points, then the shift in the peak would be 
detectable over a period shorter by $\sim\sqrt{N}$. Our simulation implies 
that it would be possible to measure the orbital decay by gravitational wave 
emission through the shift of the peak in the power spectrum.

Figure~\ref{fig:lc}C shows the dependence on the inclination angle, $\theta$, of 
the amplitude of relativistic doppler boosting relative to the amplitude of the original 
bolometric light variation emitted from the secondary black hole. The black line 
represents $L_2/L_{o,2}$ (see equation~\ref{eq:blc}), whereas the dashed, dash-dotted, 
and dotted lines represent the amplitudes corresponding to $3\sigma$, $2\sigma$, 
and $1\sigma$ of the fluctuating $L_{o,2}$, respectively. The panel indicates that the 
periodic signal corresponds to 3$\sigma$ detection if $\theta\gtrsim0.21\pi$.

%
%
\subsection{Shift in the binary orbital period.}
%
The feasibility of detecting the shift in the peak of the power 
spectrum depends primarily on the semi-major axis of the binary at the first TDE. 
We consider a circular binary whose coalescence timescale 
due to gravitational wave emission is given by \cite{p64},
\begin{eqnarray}
t_{\rm{gw}}
=
\frac{5}{8}
\frac{(1+q)^2}{q}
\frac{r_{\rm{S}}}{c}
\left(\frac{a}{{\it r}_{\rm{S}}}\right)^4
=
\frac{5}{128}
\frac{1}{\pi^{8/3}}
\frac{(1+q)^2}{q}
\left(\frac{r_{\rm{S}}}{c}\right)^{5/3}
P_{\rm orb}^{8/3}
\label{eq:tgw}
\end{eqnarray}
Setting $a_0$ and $a$ as the semi-major axes of the binary 
at the first and second TDEs, respectively, the time difference 
between two TDEs, $\Delta{t}_{\rm gw}$, needs to satisfy 
the condition that the normalized frequency is significantly increased, namely  
\begin{equation}
\left(1-\frac{\Delta{t}_{\rm gw}}{t_{\rm{gw}}(a_0)}\right)^{-3/8}\ge\frac{\Omega_{\rm{orb}}}{\Omega_{\rm{orb},0}}
\label{eq:freqcond}
\end{equation}
Thus, the upper limit on a value of semi-major axis $a_0$ is,
\begin{eqnarray}
\frac{a_0}{r_{\rm {S}}}\le\left[
\frac{8}{5}
\frac{q}{(1+q)^2}
\frac{c}{r_{\rm{S}}}
\frac{\Delta{t}_{\rm gw}}{1-(\Omega_{\rm orb}/\Omega_{\rm{orb},0})^{-8/3}}
\right] ^{1/4}.
\label{eq:ac}
\end{eqnarray}

Figure~\ref{fig:a0freq}A shows the dependence of $a_0$ on $\Omega_{\rm orb}/\Omega_{\rm{orb},0}$. 
The blue and red solid lines denote $a_{\rm 0}$ if $\Delta{t}_{\rm gw}$ is given by 10 and 50 years, 
respectively. The black dashed line shows the tidal disruption radius $r_{\rm t}$ below which 
the binary can be regarded as a single black hole (see equation~\ref{eq:rt}). The range of the 
semi-major axis at the first TDE, which is detectable, is shown by the shaded area.

%
\section*{Summary and Discussion}
\label{sec:con}
%
We have simulated the tidal disruption of a star by a circular SMBH binary 
with a $1:10$ mass ratio. Our simulation indicates that the mass accretion 
rate of the secondary black hole is higher than that of the primary black hole. 
The resulting total bolometric luminosity shows a strong relativistic beaming 
effect due to the enhanced line-of-sight velocity of the secondary black hole. 
Therefore, the power spectrum naturally shows a sharp peak at the orbital 
frequency (see Figure~\ref{fig:lc}B). 

%
\subsection{Interaction with a circumbinary disk.}
%

If the SMBH binaries are located in a gas-rich environment, 
a circumbinary disk can be the agent supplying the stars. 
Accretion disks around SMBHs are predicted to be gravitationally unstable 
at large radii where they fragment into stars or planets owing to their 
self-gravity \cite{pz78,ga01,ns07}.
Similarly, a circumbinary disk could form stars in its outskirts. Hence, 
a self-gravitating circumbinary disk would naturally lead to be an enhanced TDE rate \cite{pau13}. In addition, a circumbinary disk could grind the orbits of background stars and trigger additional TDEs \cite{scr91,jj05}. Through both mechanisms, stars would migrate to the inner edge of the circumbinary disk, and subsequently approach both black holes through the Lagrange points (L2 or L3) of the binary.

When $t_{\rm{gw}}$ is shorter than the viscous timescale evaluated 
at the inner edge of the circumbinary disk, the binary is viscously 
decoupled from the circumbinary disk \cite{me05}.
The viscous timescale measured at the inner edge of the circumbinary disk is given by 
$t_{\rm vis}=r^2/\nu$ with the inner radius being $\sim2a$ \cite{al94}, where the disk viscosity $\nu=\alpha_{\rm{SS}}c_{\rm s}{H}$, expressed in terms of the sound speed $c_{\rm s}$ and the Shakura-Sunyaev viscosity parameter $\alpha_{\rm SS}$. Based on the assumption that the circumbinary disk is a standard disk, the decoupling radius is obtained by equating $t_{\rm vis}$ to $t_{\rm gw}$, giving
\begin{eqnarray}
\frac{a_{\rm{dec}}}{r_{\rm{S}}}
&=&
\left[
\frac{2^{14/5}5^{4/5}\sqrt{2}}{5}
\left(
\frac{c}{c_{\rm{s,0}}}
\right)^2
\frac{q}{(1+q)^2}
\right]^{5/13}
\left(\frac{0.1}{\alpha_{\rm SS}}\right)^{4/5}
\left(\frac{M_{\rm BH}}{M_\odot}\right)^{1/13}
\left(\frac{\dot{M}}{\epsilon\dot{M}_{\rm Edd}}\right)^{-2/13}
,
\label{eq:adec}
\end{eqnarray}
where $c_{\rm s,0}\sim1.7\times10^8\,{\rm cm\,s^{-1}}$ is 
the sound speed for typical parameters. Figure~\ref{fig:a0freq}B 
shows the black-hole mass dependence of the decoupling radius. 
The decoupling radius is clearly larger than the fiducial semi-major axis 
$a_{\rm fid}=100r_{\rm S}$ of our simulation for 
$10^5M_\odot\le{M_{\rm BH}}\le10^8M_\odot$.

Recent numerical simulations show that the rate of gas accretion from the 
circumbinary disk is modulated by the orbital period \cite{mm08,dza13}. Detection of the orbital 
decay based on the shift in the power spectrum of the light curves can also, 
in principle, be inferred in such a system. However, if the semi-major axis of the binary 
is larger than the decoupling radius, the shift of power spectrum is undetectable because 
it is too small to be identified observationally ($\Omega_{\rm orb}/\Omega_{\rm orb,0}<1.01$ 
in Figure~\ref{fig:a0freq}A). This implies that only periodic signals caused by tidal disruption events 
around the SMBH binaries with the semi-major axis shorter than the decoupling radius can provide 
a detectable shift in the power spectrum.

The TDE flares are attributed to super-Eddington accretion phases 
of short duration \cite{rees88,ek89}. They differ from a sub-Eddington accretion mode 
which may last much longer, as illustrated in Figure 3 of Farris et al. 
(2015) \cite{farris15}. The surface density of the stellar debris in our 
simulations is many orders of magnitude higher than that of a sub-Eddington 
accretion flow. We therefore expect that any pre-existing accretion disk 
would not affect significantly the dynamics of the debris.

%
The TDE rate for a binary SMBH system is dictated by the supply rate 
of stars from the outer region of the circumbinary disk to the inner cavity. 
The formation rate of the solar mass stars is estimated to be of order 
$0.01-20\,M_\odot{\rm yr^{-1}}$ in the self-gravitating accretion disk \cite{ns07}. 
If the stars can induce gap formation in the disk, they would migrate on a timescale 
that cannot be shorter than the Eddington accretion timescale as they are coupled 
to the viscous evolution of the disk. In this case, the expected TDE rate ranges 
between $0.01\,\rm{yr^{-1}}$ and the Eddington rate, 
$\dot{M}_{\rm Edd}/M_\odot\sim0.2\,\rm{yr^{-1}}(M_{\rm BH}/10^6\,M_\odot)(0.1/\epsilon)$, 
where $\epsilon$ is the mass-to-radiation conversion efficiency. On the other hand, 
the rate at which the orbits of disk-crossing stars are brought into the inner cavity 
could also be controlled by the hydrodynamic drag force acting on the trapping 
stars \cite{DNC94,jj05}.  
In this case, stars of a solar mass should migrate to the inner cavity on the radial drift timescale 
$\tau_{\rm drift}\sim4\times10^4\,\rm{yr}\,(M_{\rm bh}/10^6M_\odot)^{3/2}\left(R_{\rm d}/0.01\,\rm{pc}\right)^{3/2}(M_{\rm d}/10M_{\rm bh})^{-1}([H/R_{\rm d}]/0.01)$ for the relatively massive, geometrically thin disk, where $R_{\rm d}$ and $H$ are the disk radius and scale-height \cite{DNC94}.
Therefore, the expected TDE rate is suppressed by the radial drift timescale down to $\sim2.5\times10^{-5}\,\rm{yr^{-1}}\,(M_{\rm bh}/10^6M_\odot)^{-3/2}\left(R_{\rm d}/0.01\,\rm{pc}\right)^{-3/2}(M_{\rm d}/10M_{\rm bh})([H/R_{\rm d}]/0.01)^{-1}$. 


%
\subsection{Detectability of the events.}
%

The light curve of the radiation emitted by the jet, whose flux is 
proportional to the mass accretion rate obtained in our simulation, 
would make it possible to distinguish a binary TDE from the usual 
emission in AGNs based on the following characteristics: 
1. the angle between the line of sight and jet direction should change 
transiently for a TDE, even if there is steady feeding 
of the black hole. 2. the jetted emission shows clear periodicity 
during several tens of binary orbital periods.

\begin{figure*}
\begin{center}
\includegraphics[width=11cm]{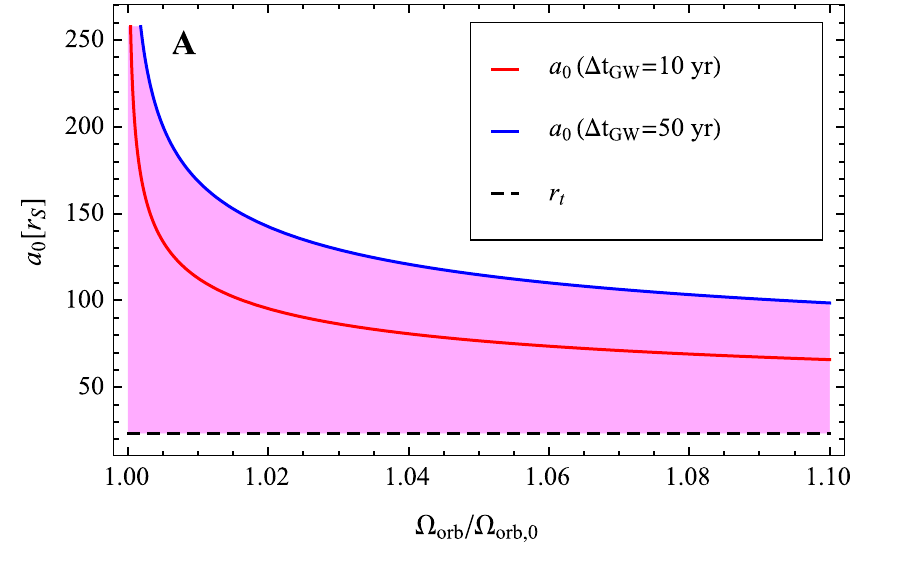}
\includegraphics[width=11cm]{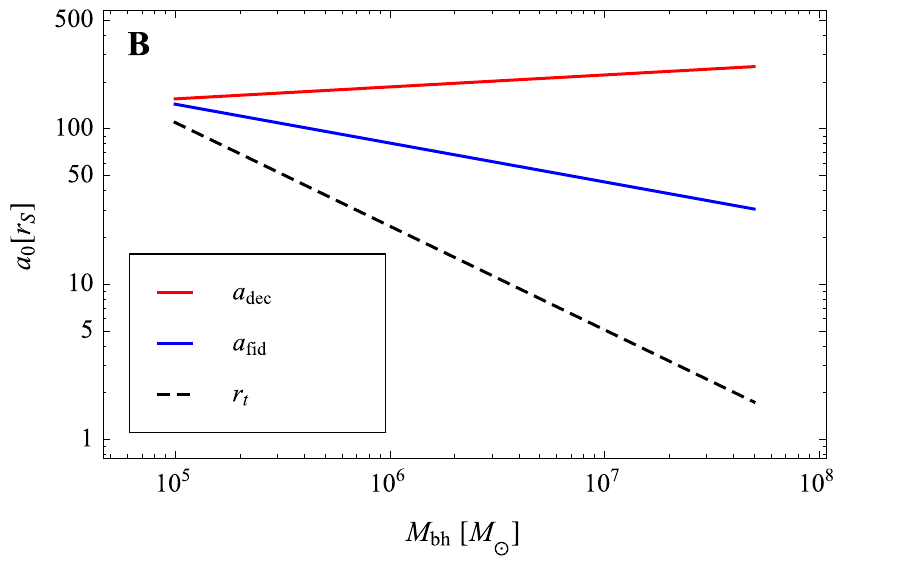}
\end{center}
\caption{
(\textbf{A}) 
Dependence of the semi-major axis at the first TDE, $a_0$, 
on the orbital frequency at the second TDE, normalized 
by the orbital frequency at the first TDE, for $M_{\rm BH}=10^{6}M_\odot$.
The blue and red solid lines show $a_{\rm 0}$ if the time difference 
between the first and second TDEs is given by 10 and 50 years, 
respectively (see equation~\ref{eq:ac}). 
The black dashed line shows the tidal disruption radius $r_{\rm t}$ below which 
the binary can be regarded as a single black hole 
(see equation~\ref{eq:rt}).
(\textbf{B}) Dependence of the decoupling radius on black hole mass.
The red, blue and black dashed lines show the decoupling radius, 
fiducial semi-major axis, and tidal disruption radius, respectively.
}
\label{fig:a0freq} 
\end{figure*}

LSST would be ideally suited for identifying candidate source for 
follow-up monitoring that will seek our predicted periodic signal. 
Given that a typical galaxy undergoes a merger 
with a mass ratio 1:10 once every $\sim1$ Gyr \cite{wcw09}, the probability for finding a 
corresponding SMBH binary in the last century of its life before 
coalescence is $\sim(10^2\,{\rm yr}/10^9\,{\rm yr})\sim10^{-7}$. 
Since the LSST will survey $3.7\times10^{10}$ galaxies for variability during its 
10 year operation (http://www.lsst.org), 
it is expected to find $\sim3700\,(\dot{N}_{\rm merger}/10^{-9}\,{\rm yr^{-1}})\,
(\dot{N}_{\rm TDE}/0.1\,{\rm yr^{-1}})(t_{\rm obs}/10\,{\rm yr})$ events.
In order to obtain the 3 sigma detection of the periodicity by relativistic 
doppler boosting, we multiply the above number by the factor 
$\int_{0.21\pi}^{0.5\pi}dS/\int_{0.0}^{0.5\pi}dS\sim0.8$, where $dS=2\pi\sin\theta{d\theta}$ 
is the differential solid angle for $\theta$ direction, from Figure~\ref{fig:a0freq}C. 
Therefore, we expect to detect $\sim2960\,(\dot{N}_{\rm merger}/10^{-9}\,{\rm yr^{-1}})\,
(\dot{N}_{\rm TDE}/0.1\,{\rm yr^{-1}})(t_{\rm obs}/10\,{\rm yr})$ 
signals from the SMBH binaries with a detectable shift over a period of $10\,{\rm year}$.


%


\begin{addendum}
\item The authors thank to the anonymous referee for fruitful 
comments and suggestions. The numerical simulations were 
performed using a high performance computing cluster at the 
Korea Astronomy and Space Science Institute. This work was 
supported in part by the research grant of the Chungbuk National 
University in 2015 [K.H.] and by NSF grant AST-1312034 [A.L.].
\item[Competing Interests] The authors have no competing financial interests.
\item[Author contributions] Both authors originated the idea for the project and worked out collaboratively its general details. K.H. performed the SPH simulation described in the Method. 
All authors reviewed the manuscript.
\item[Correspondence] Correspondence and requests for materials should be addressed to K.H.~(email: kimi@cbnu.ac.kr).
\end{addendum}


\end{document}